\documentclass[12pt, prd, aps, superscriptaddress, notitlepage]{revtex4-2}

\usepackage{amsmath}
\usepackage{graphicx}
\usepackage{mathrsfs}
\usepackage{hyperref}
\usepackage{xcolor}
\usepackage{subfigure}
\usepackage{slashed}
\usepackage{epsfig,color}
\usepackage{multirow}
\usepackage{commath}

\usepackage{float}

\renewcommand{\Re}{\text{Re}}
\renewcommand{\Im}{\text{Im}}

\begin{document}
\title{Isospin breaking corrections to a lattice QCD calculation of $\varepsilon'$}

\newcommand\bnl{Brookhaven National Laboratory, Upton, NY 11973, USA}
\newcommand\cu{Physics Department, Columbia University, New York, NY 10027, USA}
\newcommand\jlab{Thomas Jefferson National Accelerator Facility, Newport News, Virginia, USA.}
\newcommand\kntky{Department of Physics and Astronomy, University of Kentucky, Lexington, KY 40506, USA}
\newcommand\riken{RIKEN-BNL Research Center, Brookhaven National
Laboratory, Upton, NY 11973, USA}
\newcommand\soton{School of Physics and Astronomy, University of
Southampton,  Southampton SO17 1BJ, UK}

\author{Norman H. Christ}\affiliation{\cu}
\author{Erik Lundstrum}
\email{ell2156@columbia.edu}
\affiliation{\cu}

\date{June 21, 2026}

\begin{abstract}
Because of the $\Delta I = 1/2$ rule, the effects of electromagnetism and the isospin-breaking light quark mass difference on the direct CP violation parameter $\varepsilon'$ may be as large as 25\% and are consequently of immediate interest. In a lattice QCD calculation the effects of isospin breaking on the various features of kaon decay can be clearly distinguished and those effects enhanced by the $\Delta I=1/2$ rule on $\varepsilon'$ explicitly identified.  We show that all such enhanced effects can be captured in a QCD + QED lattice calculation in which the exchanged photon has an energy in an accessible, intermediate range between 0.5-2.0 GeV.  Short-distance effects ($2.0 \mathrm{\ GeV} \lesssim E_\gamma$), usually treated in QCD and electroweak perturbation theory, are not enhanced by the $\Delta I=1/2$ rule, beyond the well-understood contribution of the two electroweak penguin operators.  Infrared photons do not contribute to $\varepsilon'$ while low-energy photons ($E_\gamma \lesssim 0.5$ GeV) are not $\Delta I=1/2$ rule enhanced or are suppressed by one order in chiral perturbation theory (ChPT).  An explicit ChPT estimate of this low-energy-photon contribution, a contribution that is difficult to determine in a finite-volume lattice calculation, suggests that the effect on $\varepsilon'$ is on the order of 0.5\%.
\end{abstract}

\maketitle
\newpage
\tableofcontents

\section{Introduction}
\label{sec:intro}

Lattice QCD calculations of $\varepsilon'$, the measure of direct CP violation in $K\rightarrow\pi\pi$ decays, have reached a precision where the systematic error resulting from the neglect of the isospin-breaking (IB) effects arising from $m_u - m_d \neq 0$ and $\alpha_{\text{EM}}$ constitute a large portion of the total error budget \cite{Blum:2015ywa, PhysRevLett.115.212001, PhysRevD.102.054509, PhysRevD.108.094517}. While generally small, IB effects enter $\varepsilon'$ with a large amplification because of the $\Delta I = 1/2$ rule. Consequently, a first-principles lattice QCD calculation of these effects has become important. 

Estimates of these effects using ChPT and the large-$N$ approximation~\cite{Cirigliano:2000zw, Cirigliano_2004, Gisbert_2018, Cirigliano:2019cpi} give results as large as 25\% of $\varepsilon'$~\cite{RBC:2020kdj}.  Since these estimates rely on experimental values for the various $K\to\pi\pi$ decay rates, an estimate of the isospin-breaking contributions to $\varepsilon'$ must be part of a larger calculation of all isospin-breaking corrections to these kaon decays.

In contrast, a first-principles lattice calculation of $\epsilon'$ begins with a theoretically well-defined isospin-symmetric version of QCD to which isospin-breaking effects can be subsequently added. Since CP violation in $K\to\pi\pi$ decay requires two amplitudes with different CP-violating phases which enter the $\pi^+\pi^-$ and $\pi^0\pi^0$ decay amplitudes independently, most isospin-breaking effects do not affect $\varepsilon'$ or have an effect that appears only at higher order in ChPT when many different operators enter the calculation. Here, we exploit this fact to identify the limited group of phenomena that must be included in a lattice QCD + QED calculation of the isospin-breaking corrections to $\varepsilon'$ which are enhanced by the $\Delta I=1/2$ rule.

For example, if we are correcting an isospin-symmetric lattice QCD result for $\varepsilon'$ there are no infrared divergent corrections.  Such structure-independent corrections clearly cancel in the ratios that define $\eta_{+-}$ and $\eta_{00}$, whose difference defines $\varepsilon'$.  Similarly, since most of the suppression of the $\Delta I=3/2$ amplitude arises from QCD effects at an energy scale below the high-energy, perturbative regime, electromagnetic (EM) contributions from this high-energy region will not be enhanced by the $\Delta I= 1/2$ rule.  Here we are not discussing the contributions from the electroweak penguin operators.  These EM corrections have such an important effect on $\varepsilon'$ that they are routinely included in a lattice calculation of $\varepsilon'$ and need not be discussed here.

An important focus of this paper is to estimate the contribution of low-energy photons to $\varepsilon'$. Present lattice QCD calculations of $\varepsilon'$ rely critically on the method of L\"uscher~\cite{Luscher:1990ux} and Lellouch-L\"uscher~\cite{Lellouch:2000pv} to obtain physical infinite-volume results from calculations that involve finite-volume two-body $\pi\pi$ states.  Including electromagnetism introduces new finite-volume effects that cannot be dealt with using these earlier methods. When the final state contains only a single hadron, the method of infinite-volume reconstruction~\cite{Feng:2018qpx} can be used to remove power-law finite-volume corrections.  However, the EM corrections to $\varepsilon'$ involve three-body finite-volume effects arising from $\pi\pi\gamma$ intermediate states whose finite-volume effects are an area of active research~\cite{Tuo:2024bhm}.

A partial resolution to this problem was proposed in Ref.~\cite{PhysRevD.106.014508}, where it was demonstrated that when quantizing the EM field in Coulomb gauge, the instantaneous Coulomb potential may be included in a lattice QCD calculation with exponentially-suppressed finite-volume effects.  For the remaining portion of the EM effects arising from the transverse-polarized photons, those intermediate, $\pi\pi\gamma$ states where the photon carries an energy comparable to or greater than the kaon mass will decay rapidly in Euclidean space and can be properly included in a lattice calculation without inducing power-law-suppressed finite-volume errors. This leaves only those three-body states containing low-energy transverse photons without an established treatment.

Here we demonstrate that the contribution of these problematic states is suppressed by one order in ChPT and is on the order of $0.5\%$ of $\varepsilon'$, implying that they may be excluded from the calculation without significantly affecting the final result. Thus, even if our current inability to treat these low-energy, three-body states prevents accurate calculations of the EM corrections to the $\pi^+\pi^-$ and $\pi^0\pi^0$ kaon decay amplitudes separately, the problematic portions of those calculations cancel in the combination of amplitudes which enter $\varepsilon'$. This observation opens the door for a calculation of $\varepsilon'$ which includes the most important IB effects, allowing a significant improvement in precision  over the current estimate. This strategy of ignoring the problematic $\pi\pi\gamma$ states was originally proposed in the context of calculating the long-distance contribution to $K_L \rightarrow \mu^+\mu^-$~\cite{PhysRevD.110.054514, chao2024twophotoncontributionkltomumudecay, Boyle:2025fug}.

The same arguments which establish the small contribution of these low-energy $\pi\pi\gamma$ states also imply that the long-distance component of the Coulomb potential whose analytic treatment was described in Ref.~\cite{PhysRevD.106.014508} can also be neglected.  Thus, we adopt a more flexible, gauge-invariant, Lorentz-noncovariant approach in which all EM contributions with spatial momenta below $\mu_\mathrm{LE} \approx 500$ MeV are neglected. 

As suggested by the above discussion, two energy scales are important.  The first, $\mu_\mathrm{SD} \sim 2-3$ GeV, separates the short-distance region in which QCD perturbation theory is reliable and the long-distance region in which lattice QCD is needed.  The second scale, $\mu_\mathrm{LE} \sim 0.5$ GeV is special to the calculation of EM corrections.  The photon propagator is separated into two terms, the first in which the magnitude of the spatial momentum carried by the propagator lies below $\mu_\mathrm{LE}$ and the second with momentum magnitude above $\mu_\mathrm{LE}$.  The contribution of the first term is described as arising from low-energy photons.  It is the contribution from this first momentum region which we show is small.

The structure of this paper is as follows. Section \ref{KPP_pheno} gives a brief review of $K \rightarrow \pi\pi$ phenomenology, focusing on the limit in which EM effects and the light-quark mass difference are absent. In Section \ref{EM_eps_prime}, we generalize the usual formula for $\varepsilon'$ to include isospin-breaking effects to first order in $\alpha_\text{EM}$ and the light-quark mass difference.  We discuss the conditions, to a large degree the result of CPT invariance, that must be met if a CP-violating term in the $K\to\pi\pi$ amplitude is to contribute to $\varepsilon'$. Section~\ref{sec:ShortDistance} summarizes the treatment of energies above $\mu_{SD}$ and the effects on $\varepsilon'$ that arise when EM corrections are included at this energy scale.  Section \ref{sec:chiral_Lagrangians} describes the several chiral Lagrangians which enter a ChPT calculation of $K\rightarrow \pi\pi$ decay, including isospin-breaking effects. In Section \ref{sec:results}, we use ChPT to estimate the problematic contributions from the $\pi\pi\gamma$ states where, in the kaon rest frame, the photon has a momentum below $\mu_\mathrm{LE}$.  We use the results of Refs.~\cite{Cirigliano:2000zw,  Cirigliano_2004, Gisbert_2018, Cirigliano:2019cpi} to perform a ChPT estimate showing that these contributions are on the order of $0.5\%$ of $\varepsilon'$.  Conclusions and further discussion are given in Sec.~\ref{conclusions}.

\section{$K \rightarrow \pi\pi$ with isospin symmetry}
\label{KPP_pheno}

We begin with the general phenomenology of kaon decays. Consider first the s-wave $K^0\rightarrow\pi\pi$ decay excluding IB effects. Bose symmetry requires that the pions in the final state have isospin $I = 0$ or $I = 2$ and Watson's theorem, which is based on time-reversal symmetry, can be used to determine the complex phases associated with the amplitudes, resulting in the expressions for the $K^0 \rightarrow \pi^+\pi^-$ and $K^0 \rightarrow \pi^0\pi^0$ decay amplitudes:
\begin{equation}
\begin{aligned}
    & A_{+-} = \sqrt{\frac{2}{3}} \Big[ A_2 e^{i\delta_2}  + \sqrt{2} A_0 e^{i\delta_0} \Big] \\
     & A_{00} = \sqrt{\frac{2}{3}} \Big[ \sqrt{2} A_2 e^{i\delta_2} - A_0 e^{i\delta_0} \Big]. \\
\end{aligned}
\label{k_decay_charged_basis}
\end{equation}
Here, $\delta_I$ are the strong $\pi\pi$ s-wave scattering phase shifts in the isospin channels $I = 0$ and 2. The amplitudes $A_0$ and $A_2$ in Eq.~\eqref{k_decay_charged_basis} result from an isospin decomposition of $\pi^+\pi^-$ and $\pi^0\pi^0$ decay amplitudes into those for $\pi\pi$ states with definite isospin.  Our amplitudes are normalized so that the partial $K^0$ decay rates are given by:
\begin{eqnarray}
\Gamma_{K^0 \rightarrow \pi^+\pi^-}
 &=&\frac{1}{8\pi}|A_{+-}|^2\frac{\sqrt{m_K^2/4-m_\pi^2}}{m_K^2} \label{eq:+-rate} \\
\Gamma_{K^0 \rightarrow \pi^0\pi^0}
 &=&\frac{1}{8\pi}|A_{00}|^2\frac{\sqrt{m_K^2/4-m_\pi^2}}{m_K^2}. \label{eq:00rate}
\end{eqnarray}
In the limit of CP conservation, the CPT theorem implies symmetry under the time-reversal operation $T$ so that Watson's theorem is valid and $A_0$ and $A_2$ are purely real; CP violation allows $A_0$ and $A_2$ to take on complex values. 

The violation of CP symmetry in $K\to\pi\pi$ decay allows the long-lived neutral $K$ meson, $K_L$, to decay to two pions and the experimentally determined ratios:  $\eta_{+-}$ and $\eta_{00}$:
\begin{equation}
    \eta_{+-} = \frac{\langle \pi^+\pi^- | H_W | K_L \rangle}{\langle \pi^+\pi^- | H_W | K_S \rangle}
\label{eta+-}
\end{equation}
\begin{equation}
    \eta_{00} = \frac{\langle \pi^0\pi^0 | H_W | K_L \rangle}{\langle \pi^0\pi^0 | H_W | K_S \rangle},
\label{eta00}
\end{equation}
characterize these two CP-violating decays, where $H_W$ is the $\Delta S=\pm1$ effective weak Hamiltonian.  Indirect CP violation, arising from the decay eigenstate $K_L$ being a mixture of CP even and CP odd states, enters $\eta_{+-}$ and $\eta_{00}$ in the same way.  Therefore their difference, defined as $\varepsilon'$,
\begin{equation}
    \varepsilon' = \frac{1}{3} \left( \eta_{+-} - \eta_{00} \right).
\label{eps_eta}
\end{equation}
measures direct CP violation and results from the difference in the phases of the amplitudes $A_0$ and $A_2$, as is discussed in greater depth in Sec.~\ref{EM_eps_prime}.

We can use Eq.~\eqref{k_decay_charged_basis}, \eqref{eta+-}, \eqref{eta00} and \eqref{eps_eta} to write $\varepsilon'$ in terms of the amplitudes $A_0$ and $A_2$.  It is customary to expand the result to first order in the small ratio Re$A_2$/Re$A_0\approx 1/22$ and obtain the standard formula
\begin{equation}
    \varepsilon' = \frac{i}{\sqrt{2}} e^{i(\delta_2 - \delta_0)}\frac{\text{Re}A_2}{\text{Re}A_0} \bigg[ \frac{\text{Im}A_2}{\text{Re}A_2} - \frac{\text{Im}A_0}{\text{Re}A_0} \bigg].
    \label{eq:epsp-LO}
\end{equation}

In a lattice QCD calculation where the first-principles calculation of these quantities is possible, it is convenient to define an isospin-symmetric world in which the effects of EM and the light quark mass difference are omitted.  Such an unphysical world can be concretely specified by adopting a convenient physical quantity, often the mass of the $\Omega^-$ baryon in MeV, to determine the lattice spacing.  Two additional particle masses, for example the mass of the $(\pi^\pm, \pi^0)$ pion triplet and the mass of the two degenerate kaon doublets $(K^+,K^0)$ and $(\overline{K}^0,K^-)$, might then be required to agree with the values in MeV of the physical $\pi^0$ and $K^0$ particles, in order to determine the light and strange quark masses.  The resulting unphysical, isospin-symmetric world will be close to the physical world, differing by the $O(1\%)$ isospin-breaking effects that are being neglected. 

One then includes the effects of EM and the difference in masses of the up and down quarks to first order in perturbation theory by making first-order changes in the assigned lattice spacing and the strange, up and down quark masses to achieve first-order agreement with the physical masses of the $\Omega^-$, $\pi^0$, $K^+$ and $K^0$.  After these perturbative corrections have been made the resulting theory should disagree with nature only through effects that are second-order in isospin breaking.  Adopting this point of view substantially simplifies the calculation of the isospin-breaking contributions to $\varepsilon'$, especially if we are interested in determining those corrections to an accuracy of only 10\%.  

We will adopt this point of view and interpret the amplitudes $A_0$ and $A_2$ and Eqs.~\eqref{k_decay_charged_basis}, \eqref{eq:+-rate}, \eqref{eq:00rate} and \eqref{eq:epsp-LO} as precisely describing $K\to\pi\pi$ decay in the absence of isospin breaking and view the effects of isospin breaking described in the next section as arising from four additional amplitudes, $B^{e/o}_{00/+-}$.  Here the superscripts $e$ and $o$ indicate amplitudes that are CP-conserving or CP-violating amplitudes while $00$ and $+-$ distinguish the isospin-breaking terms in the $\pi^0\pi^0$ and $\pi^+\pi^-$ decay amplitudes.  

In contrast to what is common in the ChPT literature, our amplitudes $A_0$ and $A_2$ are not obtained from experiment and do not include isospin-breaking effects.  They can be obtained from direct calculation in an isospin-symmetric world of the sort described above.  They do depend on the five Wilson coefficients which are typically obtained from semileptonic decays and must also be defined in an unphysical, isospin-symmetric world and will also require isospin-breaking corrections.  However, we treat Eqs.~\eqref{eta+-}, \eqref{eta00} and \eqref{eps_eta} as exact, including the effects of isospin breaking to all orders with the right-hand sides of Eqs.~\eqref{eta+-}, \eqref{eta00} being physical, measured quantities.  In contrast, Eq.~\eqref{eq:epsp-LO} is accurate only to zeroth order in isospin breaking and it is the corrections to this formula that we would like to identify and then compute.

\section{Isospin-breaking contributions to  $\varepsilon^\prime$}
\label{EM_eps_prime}

In this section we add isospin-breaking effects into the determination of $\varepsilon'$ and derive a generalization of the standard Eq.~\eqref{eq:epsp-LO} for $\varepsilon'$ which includes these effects to first order.  We will include the combined effects of the $\Delta S = 1$ effective weak four-quark Lagrangian 
\begin{equation}
    H_W = H_W^e + H_W^o
\end{equation}
where $H_W^e$ and $H_W^o$ are the CP-conserving and CP-violating components of $H_W$, and the first-order effects of EM and the light quark mass differences. We define the second-order decay amplitudes:
\begin{align}
    iB_Y^X =  \sigma^X \bigg\langle (\pi\pi)_Y \bigg|& T \bigg\{  
    \bigg[-i \int d^4x \frac{m_d - m_u}{2} \bigl(\bar{d}d(x) - \bar{u}u(x)\bigr) \label{B_definition}  \\
     & - \frac{1}{2!} \int d^4x \int d^4y J_\mu(x)A^\mu(x) J_\nu(y)A^\nu(y) 
    \bigg] \mathcal{H}_W^X(0)\bigg\} - i\mathcal{H}^X_\textrm{EM-SD}(0) \bigg| K^0 \bigg\rangle, \nonumber
\end{align}
where $X = e$ or $o$, $Y = +- $ or $00$ and we choose $\sigma^e = 1$, and $\sigma^o = i$ to extract the CP-violating phase.  Here $\mathcal{H}^X_Z$ is the Hamiltonian density corresponding to the Hamiltonian $H^X_Z$ while the bra and ket states are zero-momentum QCD energy eigenstates.

In addition to the first and second terms on the right-hand side of Eq.~\eqref{B_definition} which represent the light quark mass difference and the second order EM interaction, a third effective four-quark operator $\mathcal{H}^X_\textrm{EM-SD}$ has been added.  
This operator represents the 1-loop EM corrections to the $O(G_F)$ Standard Model, $\Delta S = \pm 1$ weak decay that comes from photons with energy greater than $\mu_{SD}$.  In fact, with our exclusion of the four electroweak penguin operators, $\mathcal{H}_\mathrm{EM-SD}$ contains the same five four-quark operators which appear in $\mathcal{H}_W$ but with altered, $O(\alpha_\mathrm{EM})$ coefficients~\cite{Buchalla:1995vs}.  It includes counterterms needed to renormalize the lattice-regulated current-current product appearing in the second term in Eq.~\eqref{B_definition}.  Recall that previous RBC-UKQCD calculations of $A_0$ and $A_2$ already include the four-quark, electroweak penguin operators. Therefore, we exclude these operators from $\mathcal{H}^X_\textrm{EM-SD}(0)$.

When adding EM corrections it is important to consider the possible occurrence of infrared singularities associated with soft virtual or emitted photons.  The two amplitudes entering $\eta_{+-}$ contain charged particles in their final states and therefore if studied separately would require the addition of decays including soft radiation to avoid the appearance of infrared (IR) singularities.  However, as mentioned in Sec.~\ref{sec:intro}, such structure-independent semi-classical effects will cancel exactly in the ratio that defines $\eta_{+-}$ since the outgoing charged particles are the same in the numerator and denominator.  If a finite volume or photon mass were introduced to make the numerator and denominator of $\eta_{+-}$ well defined, the dependence on such an IR regulator would cancel from their ratio.

As we show in Sec.~\ref{sec:eval_IB_effects}, this cancellation of the contribution of low-energy photons applies to more than the structure-independent terms just discussed. In fact, if the virtual photon momentum is included as $O(p)$ in ChPT power counting, then to the leading order in ChPT these low-energy photons do not contribute to $\varepsilon'$, resulting in a significant suppression of their effects. 

The $B_Y^X$ amplitudes defined by Eq.~\eqref{B_definition} will be complex with phases that arise from two sources: the first is the factor of $i$ that is present in $\mathcal{L}_W^o$, which results from the CP-violating phase entering the CKM matrix in the SM Lagrangian. The second can be thought of as arising from the appearance of $i\varepsilon$ in the Feynman propagators from which these amplitudes are constructed. The first source of a factor of $i$ is entirely explicit and easy to identify. The second, arising from the Minkowski-space propagation in time of on-shell intermediate states, can also be determined.  In ChPT  these imaginary contributions can be determined using Cutkosky’s cutting rules.  In a lattice calculation, the phases arising from this second source can be determined by invoking Watson's theorem and using the connection between scattering phase shifts and finite-volume two-particle energies, energies which can be directly calculated using lattice QCD~\cite{Luscher:1990ux}.

The isospin-breaking amplitudes $B_Y^X$ defined in Eq.~\eqref{B_definition} can be added to the amplitudes $A_I$ computed without IB effects to provide an expression for $\varepsilon'$ including these first-order IB effects.  Substituting Eqs.~\eqref{k_decay_charged_basis}, \eqref{eta+-}, \eqref{eta00} and \eqref{B_definition} in Eq.~\eqref{eps_eta} we find
\begin{equation}
    \varepsilon' = \frac{1}{3}\left[\frac{ e^{i\delta_2} \text{Im} A_2 + \sqrt{2}e^{i \delta_0} \text{Im} A_0 + B^o_{+-}}{e^{i\delta_2} \text{Re} A_2 + \sqrt{2}e^{i \delta_0} \text{Re} A_0  + B^e_{+-}} - \frac{\sqrt{2} e^{i\delta_2} \text{Im} A_2 - e^{i \delta_0} \text{Im} A_0  + B^o_{00}}{\sqrt{2} e^{i\delta_2}\text{Re} A_2 - e^{i \delta_0} \text{Re} A_0 + B^e_{00}}\right].
    \label{eq:epsp-IB-ratios}
\end{equation}

This equation can be simplified by treating the ratio $\text{Re}A_2/\text{Re}A_0$ and the IB terms as small parameters and expanding the denominators in Eq.~\eqref{eq:epsp-IB-ratios}.  Expanding to first order in $\text{Re}A_2/\text{Re}A_0$ and IB effects, we find the result
\begin{eqnarray}
    \varepsilon' &=& \frac{i}{\sqrt{2}} e^{i(\delta_2 - \delta_0)}\frac{\text{Re}A_2}{\text{Re}A_0} \bigg[  \frac{\text{Im}A_2}{\text{Re}A_2} - \frac{\text{Im}A_0}{\text{Re}A_0} \label{eq:IB_breaking_expression} \\
    && \hskip 1.2 in  + \frac{e^{-i\delta_2}}{3 \text{Re}A_2 \text{Re}A_0} \bigg( 
    \text{Re}(A_0) (B^o_{+-} +\sqrt{2} B^o_{00}) - \text{Im}(A_0) (B^e_{+-} + \sqrt{2} B^e_{00}) \bigg) \bigg]. \nonumber
\end{eqnarray}
This expression reduces to the standard formula for $\varepsilon'$ when the $B^{e/o}_{+-/00}$ terms are omitted. The factor of Re$A_2$ that appears in the denominator of the third term in Eq.~\eqref{eq:IB_breaking_expression} provides the $\Delta I=1/2$ rule enhancement that gives these IB effects their current importance.  Since the $\Delta I=1/2$ rule suppression is special to the $A_2$ amplitude~\cite{RBC:2012ynq, RBC:2020kdj}, we expect that the four $B$ amplitudes will not be similarly suppressed, giving the third term in Eq.~\eqref{eq:IB_breaking_expression} a factor of 22 enhancement.

The algebraic structure of the expression within the large curved brackets in Eq.~\eqref{eq:IB_breaking_expression} anticipates to a large degree a major result of this paper.  To the extent that the amplitudes $B^e_Y+iB^o_Y$ obtain their CP-violating phase from the CP-violating phase of the larger amplitude $A_0$, the four terms in those brackets will cancel and $\varepsilon'$ will be unaffected by IB corrections with this phase.  Specifically, if that common CP-violating phase is $\phi$ then the left pair of terms will have the form $W\cos(\phi)\sin(\phi)$ while the second pair of terms on the right will contain identical factors appearing in the order $W\sin(\phi)\cos(\phi)$.   The common factor $W$ may itself be complex because of QCD or QED + QCD Watson phases but these two pairs of terms are identical and will cancel.

\section{Short-distance isospin-breaking contributions to $\varepsilon'$}
\label{sec:ShortDistance}

As is discussed above, in a lattice QCD calculation of $\varepsilon'$, one begins with the effective weak Hamiltonian $H_W$ that represents the short-distance structure of the Standard Model.  In the absence of short-distance EM corrections and in the case that the effects of the charm quark are treated as short-distance, the resulting three-flavor $H_W$ is composed of five independent four-quark operators: one operator in the $(27,1)$ representation of $SU(3)_L\otimes SU(3)_R$ and four in the $(8,1)$ representation.  These five independent operators can be combined to form six traditionally identified operators: two current-current operators $Q_1$ and $Q_2$ and four QCD penguin operators $Q_3$, $Q_4$, $Q_5$, and $Q_6$ as is reviewed for example in Ref.~\cite{RBC:2001pmy}. 

It is the single $(27,1)$ operator, $O_{(27,1)}$, that contributes to the amplitude $A_2$, the amplitude that is suppressed by the $\Delta I=1/2$ rule. This suppression arises from a product of two factors: the Wilson coefficient, $C_{(27,1)}$ multiplying $O_{(27,1)}$ when it appears in $H_W$ and the on-shell $K\to\pi\pi$ matrix element of $O_{(27,1)}$, $\langle\pi\pi|O_{(27,1})|K\rangle$.  The size of $C_{(27,1)}$ was found to be suppressed relative to the other coefficients in Refs.~\cite{Gaillard:1974nj, Altarelli:1974exa} by approximately a factor three, what is now a textbook result~\cite{Peskin:1995ev}.  The largest component of the factor of 22, $\Delta I=1/2$ rule suppression comes from the matrix element $\langle\pi\pi|O_{(27,1})|K\rangle$ where a cancellation between two types of color contractions results in an approximate factor of 10 reduction~\cite{RBC:2012ynq}.

The effects of including EM corrections to this short-distance determination of $H_W$ are twofold.  First, four electroweak penguin operators, conventionally labeled $Q_7$, $Q_8$, $Q_9$, and $Q_{10}$, are added to $H_W$. As stated previously, these operators are conventionally included in a lattice QCD calculation of $\varepsilon'$ and need not be discussed here.  Second, such EM corrections will result in $O(\alpha_\mathrm{EM})$ modifications of the Wilson coefficients of the five original operators $\{Q_i\}_{1\le i \le 5}$.  Therefore, the $\approx 3\times$ suppressed Wilson coefficient $C_{(27,1)}$ may be altered by $\approx 3 \times \alpha_\mathrm{EM}$ as the four other $(8,1)$ operators, with three times larger Wilson coefficients, are mixed with it.  This enhancement is significantly smaller than the $22\times$ enhancement that is the topic of this paper and might be treated as an enhancement that could be added to improve a lattice calculation of the EM corrections to $\varepsilon'$ once complete $O(\alpha_\mathrm{EM})$ perturbative results have been computed.

\section{Chiral Lagrangians}
\label{sec:chiral_Lagrangians}
In this section we describe those portions of the QCD, QED and $\Delta S=1$ chiral Lagrangians that we use in Sec.~\ref{sec:eval_IB_effects} for the ChPT estimate of the effects of low-energy photons.  As is conventional, we represent the pseudo scalar mesons by the $3\times3$ special unitary matrix
\begin{equation}
    U(x) = \exp \left(\frac{i}{F} \boldsymbol{\phi}\right),
\end{equation}
where the hermitian matrix $ \boldsymbol{\phi} = (1/\sqrt{2}) \sum \phi^a \lambda^a$ has the explicit form
\begin{equation}
    \frac{1}{\sqrt{2}} \sum_{a=1}^8 \phi^a \lambda^a = \begin{pmatrix}
\frac{1}{\sqrt{2}}\pi^0 + \frac{1}{\sqrt{6}}\eta & \pi^+ & K^+ \\
\pi^- & -\frac{1}{\sqrt{2}}\pi^0 + \frac{1}{\sqrt{6}}\eta & K^0 \\
K^- & \bar{K}^0 & -\frac{2}{\sqrt{6}}\eta
\end{pmatrix}.
\end{equation}
where $\{\lambda^a\}_{1 \leq a \leq 8}$ are the Gell-Mann matrices. 

In contrast with Refs.~\cite{Cirigliano:2000zw, Cirigliano_2004, Gisbert_2018,Cirigliano:2019cpi, Cirigliano:1999ie, Cirigliano:1999hj} we will refer to the ChPT expansion as an expansion in powers of momenta $p^2$ and quark masses $m_q$ but perform a separate expansion in powers of the fine structure constant $\alpha_\mathrm{EM}$.  Our somewhat awkward notation will imply that the QED gauge covariant derivative $\partial_\mu +ie A_\mu$ will contain terms of both first and zero order in ChPT.  However, this choice makes it easier to interpret the size of terms of order $\alpha_\mathrm{EM}$ where a term of first order in ChPT will actually behave as $p^2$. As a result, a specific term in a chiral Lagrangian may contain terms of order $p^2$ in our ChPT expansion and zeroth order in $\alpha_\mathrm{EM}$ but also terms of first order $\alpha_\mathrm{EM}$ which are zeroth order in ChPT.  We will use the term ``leading order'' to identify the term in the ChPT expansion which contains the fewest factors of (momenta$)^2$ and quark masses.  By the usual power-counting rules of ChPT, a meson loop will contribute one power of $p^2$.  However, adding a loop containing a photon will increase by one the order in $\alpha_\mathrm{EM}$ but will not increase the order in ChPT unless some further expansion in powers of the external momentum is required.

The chiral Lagrangians contain a number of {\it a priori} unknown operator coefficients, the low-energy constants or LECs. These must be determined either directly from experiment or from a more complete theory whose low-energy limit defines the effective field theory being studied. In what follows we make our estimates using the values of the LECs provided in Ref.~\cite{Cirigliano:2019cpi}.

The strong interaction chiral Lagrange density has the form
\begin{equation}
    \mathcal{L}_{\text{str}} = \frac{F^2}{4} \langle D_\mu U D^\mu U^\dag \rangle + \frac{F^2}{4} \langle \chi U^\dag + U \chi^\dag \rangle + O(p^4)
\label{eq:strong_chpt_lag}
\end{equation}
\noindent Here, $\langle \ldots\rangle$ denotes the trace over $SU(3)_L$ flavor indices.  While identified as the chiral Lagrangian describing the low-energy strong interactions, we will also include the EM couplings dictated by local QED gauge invariance.  Thus, the covariant derivative $D_\mu$ appearing in Eq.~\eqref{eq:strong_chpt_lag} is given by $D_\mu U= \partial_\mu U + ieA_\mu [Q, U]$, where $A_\mu$ is the photon field and $Q = \text{diag}\{2/3, -1/3, -1/3\}$ is the quark charge matrix. The second term containing $\chi \equiv 2B_0 \,\text{diag}\{m_u, m_d, m_s\}$ adds the effects of the up, down and strange quark masses and is responsible for the non-zero meson masses. The constant $F$ is the pion decay at leading order. The $O(p^4)$ terms generate higher-order corrections to the meson interactions which are suppressed at low-momenta.

To describe strangeness-changing weak transitions in ChPT we use the weak $\Delta S = 1$ effective Lagrangian density which can be written as:
\begin{equation}
\begin{aligned}
    \mathcal{L}^{\Delta S = 1}_{W} &=  G_8 F^4 \left < \lambda D^\mu U^\dag D_\mu U \right> + G_8 F^2 \sum_i N_i O_i^8  \\
    & +G_{27} F^4\left( L_{\mu23} L^\mu_{11} + \frac{2}{3} L^\mu_{21} L^\mu_{13}\right) + G_{27} F^2 \sum_i D_i O_i^{27} + O(G_F p^6),\end{aligned}
\label{eq:weak_chpt_lag}
\end{equation}
following a notation close to that of Eq.~(7) in Ref.~\cite{Cirigliano:2019cpi}.  The matrix $\lambda = (\lambda_6 -i\lambda_7)/2$ describes the desired  strangeness-changing transition and $L_\mu = iU^\dagger\partial_\mu U$.  The two operators on the right-hand side of the first line of this equation transform as $SU(3)$ octets and belong to the $(8, 1)$ representation of $SU(3)_L\otimes SU(3)_R$.  In the limit of isospin symmetry these operators induce purely $\Delta I = 1/2$ transitions.  The left-most term is of order $p^2$ while the right-most term behaves as $p^4$.  The second line of this equation contains 27-plet operators which transform in the  $(27,1)$ representation with the left-most term again of order $p^2$ and the right-most term of order $p^4$.  In the isospin-symmetric limit, these 27-plet operators contribute to both the $\Delta I = 1/2$ and $\Delta I = 3/2$ amplitudes, but induce the entirety of the $\Delta I = 3/2$ amplitude $A_2$.  The octet and 27-plet operators can be identified by their common coefficients $G_8$ and $G_{27}$ respectively. The coefficients $G_8$ and $G_{27}$ are proportional to $G_F$ and are complex in the presence of CP-violation. The $\Delta I =1/2$ rule in this context dictates that $G_8$ is much larger than $G_{27}$.  An explicit list of the relevant NLO operators $O_i^8$ and $O_i^{27}$ appearing in Eq.~\eqref{eq:weak_chpt_lag} can be found in Appendix A of Ref.~\cite{Cirigliano_2004}. 

In our calculation, we require only the standard EM Lagrangian density
\begin{equation}
    \mathcal{L}_{\text{elm}} = -\frac{1}{4}F_{\mu\nu} F^{\mu\nu},
\end{equation}
to which we must add a gauge-fixing term. The full ChPT EM Lagrangian contains terms responsible for meson EM mass shifts and higher-order terms which provide the effective field theory realization of the contribution from high-energy photons above the ChPT scale. We concern ourselves here with only the effects from low-energy photons, $E_\gamma 
\le \mu_{\mathrm(LE}$ so these EM LECs are not relevant.

Electromagnetic effects also come from the electroweak penguin operators.  These contributions transform as $(8, 8)$ under chiral rotations and are represented by a separate chiral Lagrangian with the appropriate transformation properties. Since lattice QCD calculations of $\varepsilon'$ already include the effective four-quark operators describing these interactions we need not consider these contributions to the $\Delta S=1$ effective Lagrangian in the current investigation.

\section{Contributions of low-energy photons to $\varepsilon'$ in ChPT}
\label{sec:eval_IB_effects}

In this section we apply the formulation of ChPT summarized in Section~\ref{sec:chiral_Lagrangians} together with Eq.~\eqref{eq:IB_breaking_expression} to determine what must be calculated to estimate the leading-order EM corrections to $\varepsilon'$ arising from photons with spatial momentum bounded by $\mu_\mathrm{LE} \approx M_K$.  We note that this is an ideal application of ChPT since ChPT is explicitly formulated to provide information in this energy range.  No new LECs must be introduced since only integrals of momenta in the range where ChPT is applicable need to be evaluated.

Central to our calculation of the four correction terms that appear on the left-hand side of Eq.~\eqref{eq:IB_breaking_expression} is the potential cancellations within each pair of terms that involve the same pion charges:
\begin{equation}
    \text{Re}(A_0) B^o_{Y} - \text{Im}(A_0) B^e_{Y},
\label{eq:IB_charged_pi_states}
\end{equation}
where $Y=+-$ or 00.  To the extent that a contribution to $A_0$, $|\Delta A_0|e^{i\phi}$, and a corresponding contribution to $B_Y$, $\Delta B^e_Y = \Delta B_Y\cos\phi$ and $\Delta B^o_Y = \Delta B_Y \sin\phi$ have the same CP-violating phase $\phi$, then their contributions to the right-hand side of Eq.~\eqref{eq:IB_breaking_expression} and to $\varepsilon'$ will cancel:
\begin{equation*}
\bigl[|\Delta A_0| \cos\phi\bigr] \bigl[\Delta B_Y\sin\phi\bigr] - \bigl[|\Delta A_0|\sin\phi\bigr] \bigl[\Delta B_Y \cos\phi\bigr] = 0.
\end{equation*}

We begin by writing expressions for $A_0$ and $B_Y$ as would be calculated in ChPT organized by the term in the $\Delta S=1$ effective Lagrangian from which that term came.  The complex amplitude $A_0$, evaluated to NLO in ChPT is then given by the formula:
\begin{equation}
\begin{aligned}
    A_0 & = G_8 \Big[ \mathcal{A}^{(p^2)} + \mathcal{A}^{(p^4)} \Big] 
    + G_8\sum_{i} N_i \mathcal{A}_i^{(p^4)}.  
\end{aligned}
\label{eq:A0_chpt_expression}
\end{equation}
Here the expression in square brackets multiplying the coefficient $G_8$ is the result of LO and NLO ChPT calculations using the QCD chiral Lagrangian given in Eq.~\eqref{eq:strong_chpt_lag} of the contribution from the leading-order $\Delta S = 1$ operator (the left-most term in the first line on the right-hand side of Eq.~\eqref{eq:weak_chpt_lag}) to the $I=0$, $K\to(\pi\pi)_{I=0}$ amplitude $A_0$ (without isospin breaking).  Likewise, the right-most term in Eq.~\eqref{eq:A0_chpt_expression}, involving the sum over $i$, is composed of similar, $I=0$, $K\to(\pi\pi)_{I=0}$ matrix elements of the NLO $\Delta S=1$ weak operators, which appear right-most in the first line of Eq.~\eqref{eq:weak_chpt_lag} with the coefficients $G_8N_i$.  This third group of matrix elements is evaluated to LO in QCD ChPT.  Since by definition the amplitude $A_0$ is the $K\to(\pi\pi)_{I=0}$ matrix element of the $\Delta S=1$ effective weak operator multiplied by the inverse of the Watson phase factor, this same $e^{-i\delta_0}$ factor must be included in the amplitudes $\mathcal{A}^{(p^2)}$, $\mathcal{A}^{(p^4)}$ and $\mathcal{A}_i^{(p^4)}$ and the product evaluated at the indicated order of ChPT.  

Next we write a similar expression for the EM amplitude $B_Y$.  
When calculating $B_Y$, we work in the kaon rest frame and consider those diagrams with a single virtual photon.  Although, as we have argued, the EM corrections to $\varepsilon'$ involve no semi-classical infrared effects, the individual terms which contribute to $B_Y$ contain infrared singularities which cancel when these terms are combined to form $\varepsilon'$.  Therefore, when computing the components of $B_Y$ below we impose a lower limit $\mu_{IR}$ on the photon momentum in the kaon rest system. 

In order to obtain the expected enhancement of EM effects from the $\Delta I=1/2$ rule, we are performing radiative corrections to the octet operators in ChPT -- the operators in Eq.~\eqref{eq:weak_chpt_lag} not suppressed by the $\Delta I=1/2$ rule.  Therefore, the complex CP-violating coefficients that enter the following expression are very similar to those in Eq.~\eqref{eq:A0_chpt_expression}
\begin{equation}
\begin{aligned}
    B_Y & = \alpha_{EM} G_8 \Big[ \mathcal{B}^{(p^2)}_Y + \mathcal{B}^{(p^4)}_Y \Big]
     + \alpha_{EM} \sum_i G_8N_i \mathcal{B}^{(p^4)}_{Yi}.
\end{aligned}
\label{eq:B_chpt_expression}
\end{equation}
This equation determines the amplitude $B_Y$ to NLO in chiral perturbation theory.  The two terms in the square bracket on the right-hand side of this equation are one-loop EM corrections to the $K\to(\pi\pi)_{I=0}$ matrix element of the left-most, $O(p^2)$ term on the right-hand side of the first line of Eq.~\eqref{eq:weak_chpt_lag}, computed to LO and NLO in QCD ChPT, respectively.  The right-most term in Eq.~\eqref{eq:B_chpt_expression} is the one-loop EM correction to the $K\to(\pi\pi)_{I=0}$ matrix element of the right-most, $O(p^4)$ term, also in the first line of Eq.~\eqref{eq:weak_chpt_lag} and computed to LO in QCD ChPT. As described above, the energy of the photon in these one loop corrections is bounded above by $\mu_\mathrm{LE}$ and bounded below by energy $\mu_{\mathrm{IR}}$.  Thus, the amplitudes $\mathcal{B}^{(p^2)}$, $\mathcal{B}^{(p^4)}$ and $\mathcal{B}^{(p^4)}_i$ in Eq.~\eqref{eq:B_chpt_expression} depend on these two variables.

As in Eq.~\eqref{eq:A0_chpt_expression}, in each case in Eq.~\eqref{eq:B_chpt_expression} the superscript $(p^n)$ indicates the actual order in the usual ChPT expansion of the designated term.  As discussed earlier, while adding a one-photon loop removes two powers of $p$ because the $p_\mu$ terms in both QED covariant derivatives are replaced by one end of a photon propagator, the bound $|\vec p\,| \le \mu_\mathrm{LE}$ obeyed by the spatial loop momentum $\vec p$ introduces its own factor of $\mu_\mathrm{LE}^2$ leaving the order $n$ of $p^n$ unchanged.  Finally, we should recall that in addition to the explicit factors of $G_8$ and $G_8N_i$ displayed in Eq.~\eqref{eq:B_chpt_expression} there is also a factor of the inverse of the Watson phase, $e^{-i\delta_0}$, present in the amplitudes $\mathcal{B}$ that must also be part of the QCD ChPT expansions.

We can now substitute Eqs.~\eqref{eq:A0_chpt_expression} and \eqref{eq:B_chpt_expression} into Eq.~\eqref{eq:IB_charged_pi_states} to obtain an explicit expression for the EM contribution to $\varepsilon'$: \begin{eqnarray}
\text{Re}(A_0) B^o_Y - \text{Im}(A_0) B^e_Y && 
 \label{eq:IB_Chpt_NNLO_expression}\\
  &&\hskip -1.6 in =\Big\{\text{Re}(G_8) \mathcal{A}^{(p^2)} 
    + \sum_{i}\text{Re}(G_8 N_i) \mathcal{A}_i^{(p^4)}\Big\} 
\times \Big\{\alpha_{EM} \text{Im}(G_8)  \mathcal{B}^{(p^2)}_Y 
+\alpha_{EM} \sum_i \text{Im}(G_8N_i) \mathcal{B}^{(p^4)}_{Yi} \Big\} 
 \nonumber \\
  && \hskip -1.5 in - \Big\{\text{Im}(G_8) \mathcal{A}^{(p^2)} 
    + \sum_{i}\text{Im}(G_8 N_i) \mathcal{A}_i^{(p^4)}\Big\} 
       \times \Big\{\alpha_{EM} \text{Re}(G_8)  \mathcal{B}^{(p^2)}_Y +\alpha_{EM} \sum_i \text{Re}(G_8N_{Y,i}) \mathcal{B}^{(p^4)}_{Yi} \Big\}. 
 \nonumber
\end{eqnarray}
We have simplified this expression by keeping only the leading-order ChPT amplitude for each of the distinct factors, $G_8$ and $G_8N_i$ which carry CP-violating phases.

As can be seen from Eq.~\eqref{eq:IB_Chpt_NNLO_expression} the terms that are leading-order in ChPT cancel leaving the two NLO terms:
\begin{eqnarray}
\text{Re}(A_0) B^o_Y - \text{Im}(A_0) B^e_Y &=&\alpha_{EM} \sum_i\mathcal{A}^{(p^2)}\mathcal{B}_{Yi}^{(p^4)}
      \Big[\text{Re}(G_8)\text{Im}(G_8N_i) - \text{Im}(G_8)\text{Re}(G_8N_i)\Big]  \label{eq:IB_Chpt_NLO_expression} \\
  && \hskip 0.1 in +\alpha_{EM} \sum_i\mathcal{A}_i^{(p^4)}\mathcal{B}^{(p^2)}_Y
      \Big[\text{Re}(G_8N_i)\text{Im}(G_8) - \text{Im}(G_8N_i)\text{Re}(G_8)\Big],
 \nonumber
\end{eqnarray}
where, in this expression, we have omitted terms of order $p^8$.  Equation~\eqref{eq:IB_Chpt_NLO_expression} is an important result of this paper.  The effects of the low-energy photons which are difficult to accurately include in a finite-volume lattice QCD+QED calculation of $\varepsilon'$ give a contribution that is suppressed by one order $p^2$ when computed using ChPT.  We will attempt to estimate the size of these suppressed low-energy-photon effects in the following section.

Before closing this section, we discuss two additional topics.  The first is the isospin breaking effects of the mass terms for each of the three active quark flavors needed to shift the isospin-symmetric quark masses present in the original isospin-symmetric world whose prediction for $\varepsilon'$ we are trying to correct.  These include counterterms needed to adjust the self-energy effects of the lattice photons that are cutoff by the lattice scale to those which give the physical ratios of $M_K/M_\Omega$ and $M_\pi/M_\Omega$.

Since these small mass shifts lie in the energy range in which ChPT is accurate, their effects might also be estimated using ChPT. Just as in the case of the low-energy photon contribution discussed here, because of their $O(p^2)$ size in ChPT, their effects will also be suppressed by one power of $p^2$ and might be neglected in a first calculation of the isospin breaking corrections to $\varepsilon'$. However, the effects of these mass shifts are straight-forward to determine in a lattice calculation by simply repeating the calculation using different input quark masses.

The second topic, introduced for completeness, is a comparison between the contribution of the low-energy photons discussed above with that of the two four-quark electroweak penguin operators that transform in the $(8,8)$ representation of $SU(3)_L\otimes SU(3)_R$.  To leading order these contribute to a single term in the $\Delta S=1$ ChPT effective Lagrangian density:
\begin{equation}
    \mathcal{L}_{W}^{\Delta S = 1,\text{EWP}} = \alpha_{EM} g_{\text{ewp}} F^6 \langle \lambda U^\dagger Q U \rangle,
\end{equation}
where, as in Eq.~\eqref{eq:weak_chpt_lag} the $\lambda$ matrix projects onto the $\Delta S = 1$ transition while $Q$ is the quark charge matrix. The coefficient $g_{\text{ewp}}$ is complex in the presence of CP violation with a CP-violating phase that is different from that of the complex coefficient $G_8$ in the $\Delta S = 1$ weak chiral Lagrangian. Therefore, the contribution from this operator interferes with the LO term in $A_0$ (which is $O(p^2)$ and proportional to $G_8$) and survives the difference in Eq.~\eqref{eq:IB_charged_pi_states}. Thus, the contribution of the LO electroweak penguin operator to the numerator on the right-hand side of Eq.~\eqref{eq:IB_breaking_expression} is of $O(\alpha_{EM}p^2)$.  This contribution should be substantially larger than the $O(\alpha_{EM}p^6)$ contribution of low-energy virtual photons that we propose to neglect.

\section{Estimate of low-energy photon contribution to $\varepsilon'$}
\label{sec:results}

In the previous section we established that the contribution of low-energy photons to $\varepsilon'$ is suppressed by one order of $p^2$ in chiral perturbation theory.  In this section we refine that estimate to include typical numerical factors such as powers of $4\pi$ by explicitly calculating some of the terms involving low-energy photons which contribute to the right-hand side of Eq.~\eqref{eq:IB_Chpt_NLO_expression}.  

As shown in that equation, for each possible charge assignment to the final two pions ($Y=$+- or 00) there are two types of leading-order contribution:  The first is the product of the LO factor from the isospin conserving contribution to $A_0$ multiplied by the EM correction to the contribution of NLO, isospin-symmetric $\Delta S=1$ ChPT effective Lagrangian.  This correction is given by the first line on the right-hand side of Eq.~\eqref{eq:IB_Chpt_NLO_expression}.  The second is the product of the NLO factors from the isospin-conserving contribution to $A_0$ multiplied by the EM correction to the contribution of the LO, isospin-symmetric $\Delta S=1$ ChPT effective Lagrangian.  This correction appears in the second line of Eq.~\eqref{eq:IB_Chpt_NLO_expression}.

Because of the large number of separate terms that must be computed when evaluating the first line of Eq.~\eqref{eq:IB_Chpt_NLO_expression} as described above, we choose to obtain our estimate from those on the second line. There are three noteworthy issues associated with this choice.  First, since the EM correction to the LO term in the $\Delta S=1$ chiral Lagrangian is non-zero only for the case of charged final-state pions this choice to evaluate only the second correction term implies that we need only consider the $Y=+-$ case, further simplifying our task.

Second, the cancellation of infrared divergences in the EM corrections to $\varepsilon'$ involves a cancellation of infrared divergent components between the first and second lines in Eq.~\eqref{eq:IB_Chpt_NLO_expression}.  By evaluating only the second line in that equation we have introduced an unphysical infrared divergence that would not be present in a complete calculation. We deal with this problem by introducing a lower limit $\mu_{\mathrm{IR}}$ on the magnitude of the three-momentum carried by the photon in the rest frame of the kaon when performing our calculation of the second line of Eq.~\eqref{eq:IB_Chpt_NLO_expression}. We have calculated analytically the coefficient of the IR divergent logarithm which appears in the final amplitude. The numerical value of the real part of this coefficient is smaller by a factor of three relative to the real part of the complete amplitude, and therefore we can shift the factor $\mu_{\mathrm{IR}}$ which enters the divergent logarithm by a factor of twenty before introducing a one hundred percent change in the real part of our amplitude. The imaginary part of this coefficient is only a factor of 1.5 smaller than the imaginary part of the complete amplitude. (See Eqs.~\eqref{eq:B-number} and \eqref{eq:B-log} below.) Therefore, lowering the value of $\mu_{\mathrm{IR}}$ by a factor of four will change our numerical estimate for the imaginary part at the level of one hundred percent. This is, however, not sufficient to meaningfully change the magnitude of the contribution to $\varepsilon'$, and we take this as evidence that our calculation is sufficiently insensitive to our unphysical IR limit for the purposes of this estimate and base our estimate on value $\mu_{\mathrm{IR}}=30$ MeV.

The third issue related to our evaluation of only the second line of Eq.~\eqref{eq:IB_Chpt_NLO_expression} is that the first line of this equation is also not evaluated in the work of Cirigliano {\it et al.}~\cite{Cirigliano:2019cpi}.  We expect that this omission is consistent in their calculation but it is not in ours because the EM corrections needed when $\varepsilon'$ is computed using experimentally measured quantities are of lower order in ChPT than those needed when one corrects a theoretically defined, isospin-symmetric calculation of $\varepsilon'$.

In order to make our phenomenological estimate, we require numerical values for the ChPT LECs that enter the NLO isospin-symmetric terms on the second line of Eq.~\eqref{eq:IB_Chpt_NLO_expression}. We use the LECs determined in Ref.~\cite{Cirigliano:2019cpi}.  We also use the experimental values for $\Re A_0$ and $\Re A_2$ determined from the two  $K_S\to\pi\pi$ decays assuming isospin symmetry.  Neither the LECs nor these values for $\Re A_0$ and $\Re A_2$ are not taken from an isospin-symmetric theory of $K\to\pi\pi$ decay and in our strategy for including IB effects contain unwanted IB effects.  However, these are NLO in $\alpha_\mathrm{EM}$ and $m_u-m_d$ and therefore can be consistently ignored here.

\begin{figure}[t]
  \centering
  \includegraphics[scale=0.5]{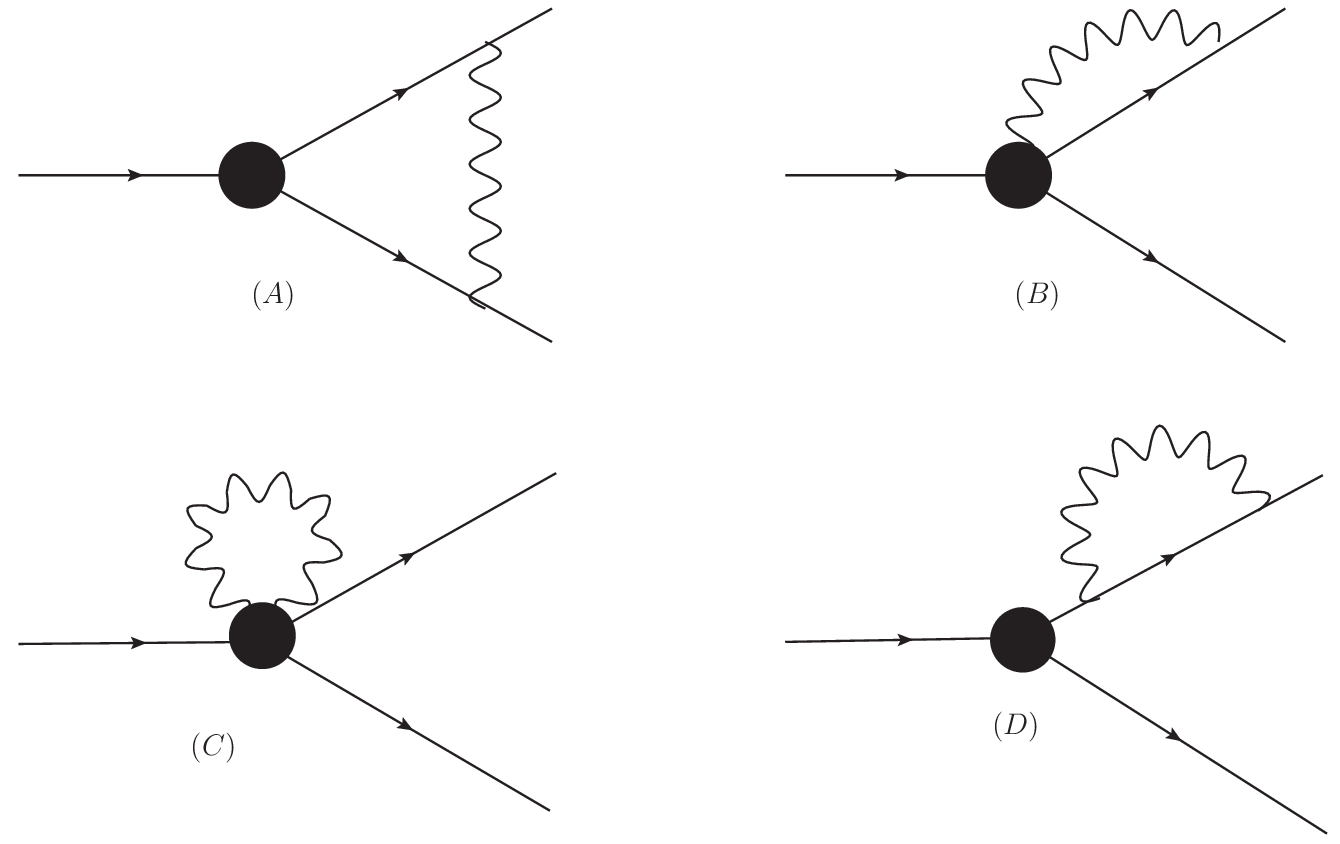} 
   \caption{Feynman diagram topologies which contribute to $B_{+-}$. The neutral kaon enters from the left and the solid black vertices are the $\Delta S = 1$ transition. Diagram D contributes through the wavefunction renormalization factor for the charged pions.}
    \label{fig:photon_diagrams}
\end{figure}

To calculate $B^{(p^2)} (\mu_{\text{LE}}, \mu_{\text{IR}})$, we quantize the EM field in Coulomb gauge and evaluate the diagrams in Fig.~\ref{fig:photon_diagrams}. To check our results we have performed the same calculation in Feynman gauge and confirmed that the two agree. If evaluated without constraints these diagrams are UV divergent and in a full ChPT calculation would require the introduction of new LECs.  Here we are interested in computing the contribution for low-energy photons with momentum $|\vec k| \le \mu_\mathrm{LE} \approx M_K$ in the rest frame of the kaon.  Thus, our result is well determined by ChPT without the need for additional information.  In order to make our estimate, we use Cauchy's theorem to evaluate the integrals over $k_0$, returning from covariant Feynman perturbation theory to ``old-fashioned'' non-covariant perturbation theory.  We then perform the integral over the photon's spatial momentum $\vec k$ within the bounds $\mu_{\mathrm{IR}}\le|\vec k| \le \mu_\text{LE}$.

In Coulomb gauge, the momentum-space Feynman propagator $D_{\mu\nu}(p)$ receives two contributions, one from the instantaneous Coulomb potential and the second from the propagating transverse-polarized photons:
\begin{equation}
    D_{00}^{\text{C}} (p) =   \frac{i}{\abs{\vec{p}}^2},
\label{coulob_potential_prop}
\end{equation}
\begin{equation}
    D_{ij}^{\text{tr}}(p) = \frac{i}{p^2 + i\varepsilon} \left( \delta_{ij} - \frac{p_i p_j}{\abs{\vec{p}}^2} \right),
\label{transverse_photon_prop}
\end{equation}
\noindent where $i,j$ take the values 1, 2 or 3. All other components of the photon propagator are zero. We use this propagator to calculate the radiative corrections to the LO $\Delta S = 1$ octet operator in Eq.~\eqref{eq:weak_chpt_lag}. For our calculation we use the pion decay constant $F = 92$ MeV, a value smaller by a factor of $\sqrt{2}$ than the usual RBC/UKQCD conventions.

In Fig.~\ref{fig:photon_diagrams} we show the types of diagram we evaluate. Diagrams such as those in Figs.~\ref{fig:photon_diagrams}(A) and ~\ref{fig:photon_diagrams}(B) explicitly contain factors $O(p^2)$ in the external momenta and conform to the power counting described above. Diagram Fig.~\ref{fig:photon_diagrams}(C) contract the photon fields in each covariant derivative, and is therefore naively $O(p^0)$. However, this diagram is quadratically divergent and therefore the final result is proportional to our UV cutoff $\mu_{\text{LE}}^2$. Since we choose $\mu_{\text{LE}}^2 \approx M_k^2$, this diagram is also $O(p^2)$.  

Our result for $\mathcal{B}^{(p^2)}_{+-}$ evaluated at $\mu_\mathrm{LE} = 500$ MeV and $\mu_\mathrm{IR} = 30$ MeV is
\begin{equation}
\mathcal{B}^{(p^2)}_{+-} = F_\pi \Big[(-3.58 + i 7.59 ) \times 10^{-1}\text{ GeV}^2 \Big]  \label{eq:B-number}
\end{equation}
while the explicit dependence on the lower limit $\mu_{\mathrm{IR}}$ discussed above corresponds to the term
\begin{equation}
    F_\pi \Big[(- 1.32  - i5.05)\times 10^{-1}  \text{ GeV}^2 \Big] \log(\mu_{\mathrm{IR}}) \label{eq:B-log}
\end{equation}
that is present in $\mathcal{B}^{(p^2)}_{+-}$. The ratio of the coefficients shown in Eqs.~\eqref{eq:B-number} and \eqref{eq:B-log} was referred to above.

The other numerical values of the quantities that enter our estimate are given in Table~\ref{value_table}, and are all taken from the values determined in Ref.~\cite{Cirigliano:2019cpi}. It should be noted that the conventions for $A_0$ and $A_2$ used by the RBC-UKQCD collaboration in Refs.~\cite{Blum:2015ywa,PhysRevLett.115.212001,PhysRevD.102.054509,PhysRevD.108.094517} differ from those used in Refs.~\cite{Cirigliano_2004,Cirigliano:2019cpi} by a factor of $\sqrt{3/2}$, which we have added to make our estimates.
The numerical values of the NLO octet amplitudes that enter Eq.~\eqref{eq:IB_Chpt_NLO_expression} can be constructed from Eqs.~(23) and (56) and Table 4 of Ref.~\cite{Cirigliano:2019cpi}. The values are
\begin{equation}
\begin{aligned}
    & \sum_i\mathcal{A}_i^{(p^4)} \text{Re}(G_8N_i) =  \sqrt{3} F_{\pi} (M_K^2 - m_\pi^2)\text{Re} G_8 \cdot \big[0.02(05)_{\chi_\text{ren}}],  \\
    & \sum_i\mathcal{A}_i^{(p^4)} \text{Im}(G_8N_i) = \sqrt{3} F_{\pi} (M_K^2 - m_\pi^2) \text{Im} G_8 \cdot \big[0.10(05)_{\chi_\text{ren}} \big].
\end{aligned}
\label{eq:NLO-IC}
\end{equation}
These quantities contain uncertainties resulting from various sources related to the matching process which determines the LECs. The matching scheme uses the large $N_c$ expansion, where $N_c$ is the number of quark colors, which misses some of the logarithmic corrections. To estimate the size of this effect, the authors of Ref~\cite{Cirigliano:2019cpi} vary the chiral renormalization scale at which the matching is performed. This becomes the dominant source of uncertainty in the calculations. 

The error associated with the chiral renormalization scale is correlated in all quantities.  However, for the purposes of our estimate it is sufficient to treat the uncertainties as uncorrelated. Other uncertainties result from the short-distance matching scale used to determine the LECs and the mass of the strange quark. All errors are propagated through separately and combined in quadrature for the final estimate.

\begin{table}
\begin{center}
\begin{tabular}{ |p{2.3cm}||p{5.8cm}|  }
 \hline
 \multicolumn{2}{|c|}{Values entering estimate} \\
 \hline
 Re$A_0$ & $3.320(1)_{\text{exp}}\cdot 10^{-7}$ GeV  \\
 \hline
 Im$A_0$ & $-3.65(21)_{\chi_\text{ren}}(38)_{\mu_\text{SD}}(42)_{L_5}(13)_{m_s} \cdot10^{-11}$ GeV  \\
 \hline
 Re$A_2$ & $1.497(4)_{\text{exp}}\cdot 10^{-8}$ GeV  \\
 \hline
  $\text{Re}G_8 $ & $ -6.48(25)_{\chi_\text{ren}}\cdot 10^{-6}$ $\text{GeV}^{-2}$\\
 \hline
  $\text{Im}G_8 $ & $ 9.32(1.07)_{\mu_\text{SD}}(1.19)_{L_5}(0.36)_{m_s} \cdot 10^{-10}$ $\text{GeV}^{-2}$\\
 \hline
\end{tabular}
\end{center}
\caption{Values entering estimate of low-energy photon contribution to $\varepsilon'$. Uncertainties are given in parentheses and result from the following sources. Those on Re$A_0$ and Re$A_2$ are experimental. Uncertainties entering $\text{Re}G_8$ and $\text{ImA}_0$ come from varying the chiral scale at which the matching to experiment is performed. As discussed in the text, the value for $\text{Im}G_8$ is determined by matching to the SM via the large $N_c$ approximation. The errors associated with $\text{Im}G_8$ are estimated from the dependence on short-distance scale at which matching is performed, uncertainties in the values of strong chiral LECs and in the mass of the strange quark, respectively.}
\label{value_table}
\end{table}

\begin{figure}[t]
  \centering
  \includegraphics[scale=0.7]{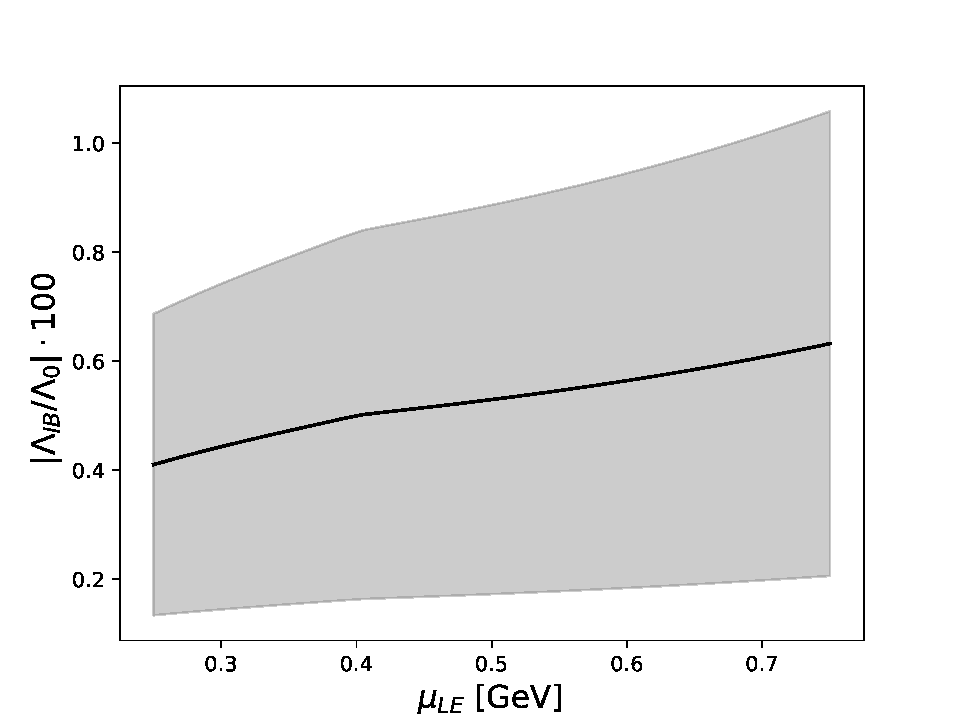} 
   \caption{The estimated relative contribution of low-energy propagating $\pi\pi\gamma$ states to $\varepsilon'$ plotted as a function of the maximum allowed photon momentum in the kaon rest frame.  As explained in the text, we also use the lower bound $\mu_{\mathrm{IR}} = 30$ MeV on the photon momentum. Plotted uncertainties, shown by the shaded region, are dominated by theoretical uncertainties in the values of chiral LECs. We observe effects on the order of 0.5\% of $\varepsilon'$.}
    \label{fig:IB-cont}
\end{figure}

To complete the estimate, we calculate the low-energy-photon, IB term in Eq.~\eqref{eq:IB_breaking_expression}
\begin{eqnarray}
    \Lambda^{\text{LE}}_{\text{IB}} &=& \frac{e^{-i\delta_2}}{3 \text{Re}A_2 \text{Re}A_0} \big[ \text{Re}(A_0) B^o_{+-} - \text{Im}(A_0) B^e_{+-} \big] \label{eq:Lamgda-LE1} \\
    &=& \frac{e^{-i\delta_2}}{3 \text{Re}A_2 \text{Re}A_0} \alpha_{EM} \mathcal{B}^{(p^2)}_{+-}
      \Big[\Big(\sum_i\mathcal{A}_i^{(p^4)}\text{Re}(G_8N_i)\Big)\text{Im}(G_8) \\ 
&& \hskip 2.0 in -\Big(\sum_i\mathcal{A}_i^{(p^4)}\text{Im}
(G_8N_i)\Big)\text{Re}(G_8)\Big], \label{eq:Lamgda-LE2}, \nonumber
\end{eqnarray}
where in Eq.~\eqref{eq:Lamgda-LE2} we have specialized to the term in the second line of Eq.~\eqref{eq:IB_Chpt_NLO_expression} which is the term being evaluated in our estimate.  We can then substitue 
the numerical values given in  Eq.~\eqref{eq:NLO-IC} and Table~\ref{value_table},    We can then compare $\Lambda^{\text{LE}}_{\text{IB}}$ with the dominant term in $\varepsilon'$ also appearing in Eq.~\eqref{eq:IB_breaking_expression}
\begin{equation}
    \Lambda_0 = \frac{\text{Im}A_0}{\text{Re}A_0},
\end{equation}
where $\Im A_0$, given in Table~\ref{value_table} is the ChPT prediction in the isospin limit using the results of Ref.~\cite{Cirigliano:2019cpi}, which can be calculated from Eqs.~(23) and (48) and the values in Table 4 of this reference. For simplicity we have omitted the $\text{Im}A_2/\text{Re}A_2$ term in Eq.~\eqref{eq:IB_breaking_expression} which is $5\times$ smaller than $\Lambda_0$.  In Fig.~\ref{fig:IB-cont}, we plot the ratio $|\Lambda^{\text{LE}}_{\text{IB}} / \Lambda_0|$ as a function of the maximum allowed photon momentum $\mu_{\mathrm{LE}}$ in the kaon rest frame. For $\mu_{\text{LE}} = 500$ MeV and $\mu_{\text{IR}} = 30$ MeV, we find
\begin{equation}
    \abs{\frac{\Lambda^{\text{LE}}_{\text{IB}}}{\Lambda_0}} = 5.3(3.6) \cdot 10^{-3},
    \label{eq:estmate}
\end{equation}
implying a correction from these problematic low-energy photons that is only several tenths of a percent of $\varepsilon'$.  

It is instructive to examine the individual factors which lead to the estimate given in Eq.~\eqref{eq:estmate}:
\begin{eqnarray}
\abs{\frac{\Lambda_{\mathrm{IB}}^{\mathrm{LE}}}{\Lambda_0}} &\approx& \frac{1}{3}\frac{\mathrm{Re}{A_0}}{\mathrm{Re}A_2} \cdot \frac{\alpha_{EM} \mathcal{B}^{(p^2)}_{+-}\mathrm{Re}G_8}{\mathrm{Re} A_0} \cdot
\frac{\sum_i\mathcal{A}_i^{(p^4)}\text{Im}
(G_8N_i)}{\mathrm{Im}A_0} \label{eq:estimate-breakdown1} \\
&\approx& \frac{1}{3}\cdot 22 \cdot 0.005 \cdot 0.09 = 0.0032  \label{eq:estimate-breakdown2}
\end{eqnarray}
where in the final factor we have used only the contribution from the second line in Eq.~\eqref{eq:NLO-IC} since it is five times larger than the first.  

Moving from left to right in Eq.~\eqref{eq:estimate-breakdown1}, the factor of 1/3 might be viewed as a Clebsch-Gordan coefficient while 22 is the important enhancement coming from the $\Delta I = 1/2$ rule.  The third factor of 0.005 is the estimate presented in this paper of the one-loop EM correction to the leading-order $K\to\pi\pi$ ChPT vertex found as the first term on the right-hand side of Eq~\eqref{eq:weak_chpt_lag}.  The final factor of 0.09 is the suppression of these CP and isospin breaking effects by one order in ChPT explained earlier in this paper.  The 0.0032 result from this approximate summary is in reasonable agreement with the more careful 5.3(3.6)$\times 10^{-3}$ result given in Eq.~\eqref{eq:estmate}

This straight-forward interpretation of Eq.~\eqref{eq:estimate-breakdown1} allows us to speculate on the size of the EM corrections coming from photons more energetic than the 500 MeV cutoff imposed here.  One might expect that the estimate given in Eq.~\eqref{eq:estimate-breakdown2} would apply to this case as well, after the 0.09 ChPT suppression factor has been removed. This suggests an IB correction to $\varepsilon'$ on the order of a few percent, possibly 5-10$\times$ smaller than the estimate of a possible 25\% IB correction to the calculation given in Ref.~\cite{RBC:2020kdj} based on the IB corrections computed in Ref.~\cite{Cirigliano:2019cpi}, a correction based on a different definition of IB effects than that adopted here.  More specifically, the most accurate results from calculation of $\epsilon'$ reported in Ref.~\cite{RBC:2020kdj} made use of the experimental values for Re$A_0$ and Re$A_2$, potentially introducing the large isospin breaking effects analyzed in Ref.~\cite{Cirigliano:2019cpi}.

\section{Conclusions}
\label{conclusions}

In this paper we have examined the contribution from low-energy $\pi\pi\gamma$ intermediate states to $\varepsilon'$. These states will be significantly distorted by finite-volume errors, potentially preventing a successful lattice QCD calculation of $\varepsilon'$ which includes IB effects. Our main goal was to assess whether the states containing a low-momentum photon intermediate can be safely omitted from a lattice QCD calculation of $\varepsilon'$ without introducing  significant errors. To make our estimate we used chiral perturbation theory worked to leading order in $\alpha_{\text{EM}}$ and $\Re A_2 / \Re A_0$. Our numerical estimate relies on the results of Cirigliano {\it et al.} including the large number of LECs entering the NLO weak interaction chiral Lagrangian. We found that all states including a photon with momentum below 500 MeV/c contribute on the order of several tenths of a percent to $\varepsilon'$ and may therefore be safely omitted from a lattice QCD calculation even at 1\% precision. This conclusion takes proper account of the potential $22\times$ $\Delta I=1/2$ rule enhancement of isospin breaking effects on $\varepsilon'$.

This and other useful conclusions arise from the following qualitative features of a lattice QCD calculation of direct CP violation in $K\to\pi\pi$ decay.
\begin{enumerate}
\item  A three-flavor lattice QCD calculation of $\varepsilon'$, including isospin breaking effects to first order in $\alpha_{\mathrm{EM}}$, can be performed with a minimum of Standard Model inputs: one dimensionful quantity such as $M_\Omega$ to set the lattice scale, the $\pi^0$, $K^+$ and $K^0$ meson masses to determine the three quark masses and seven Wilson coefficients, determined from semi-leptonic meson decay.  There is no need to include experimental results for specific $K\to\pi\pi$ decay amplitudes which themselves may involve significant infrared and Coulomb corrections.
\item The experimentally measurable quantity $\varepsilon'$ is defined in terms of ratios of $K\to\pi\pi$ decay amplitudes in which the numerator and denominator involve $\pi\pi$ states with the same electric charges ensuring that all structure-independent IR effects cancel.
\item To leading order in ChPT the weak interaction effective Lagrangian involves only two terms, with complex, CP-violating coefficients $G_8$ and $G_{27}$, where the $\Delta I = 1/2$ rule requires $G_{27} \approx G_8/22$.  The effects of low-energy photons can be computed from this Lagrangian.  However, the leading-order effects in ChPT, not suppressed by the $\Delta I=1/2$ rule, will generate new terms proportional to $G_8$.  Since these new terms have the same phase as the leading order term, their interference cannot contribute to $\varepsilon'$ making any effect of these low-energy photons on $\varepsilon'$ NLO in ChPT, substantially reducing any $\Delta I = 1/2$ rule enhancement.
\item  The same argument given above for the absence of leading-order effects of low-energy photons also applies to the effects of isospin-breaking quark masses. Although such quark mass effects are straight-forward to determine in a lattice QCD calculation, their effects on $\varepsilon'$ will also not show a $\Delta I=1/2$ rule enhancement.
\item If we leave aside the electroweak penguin operators which are already included in lattice QCD calculations of $\varepsilon'$, we can also argue that short-distance EM effects on $\varepsilon'$ will also not be $\Delta I=1/2$ rule enhanced.  The effects of electromagnetism on the five Wilson coefficients that describe the non-electroweak-penguin contributions to $K\to\pi\pi$ decay should be on the order of 1\%.  Since the $\Delta I = 1/2$ rule arises from the anomalously small matrix element of a single one of these five weak operators~\cite{RBC:2012ynq}, not a subtle cancellation among them, the effect of these short-distance EM effects will also be $\Delta I = 1/2$ rule suppressed.

\end{enumerate}

We conclude that while complex, a lattice QCD calculation of the isospin breaking effects on $\varepsilon'$ with an error of 1\% of $\varepsilon'$ are theoretically possible even when the effects of the $\Delta I=1/2$ rule are considered.  Such a calculation is further simplified because the isospin breaking of the up and down quark masses need not be considered and the EM corrections to the Standard Model Wilson coefficients need not be available.

\bibliography{ref.bib}

\end{document}